# On multiple IoT data streams processing using LoRaWAN

### Kunal Chowdhury

**Abstract**— LoRaWAN has turned out to be one of the most successful frameworks in IoT devices. Real world scenarios demand the use of such networks along with a robust stream processing application layer. To maintain the exactly once processing semantics one must ensure that we have proper ways to proactively detect message drops and handle the same. An important use case where stream processing plays a crucial role is joining various data streams that are transmitted via gateways connected to edge devices which are related to each other as part of some common business requirement. LoRaWAN supports connectivity to multiple gateways for edge devices and by virtue of its different device classes, the network can send and receive messages in an effective way that conserves battery power as well as network bandwidth. Rather than relying on explicit acknowledgements for the transmitted messages we take the advantage of these characteristics of the devices to detect , handle missing messages and finally process them.

**Index Terms**— Communication/Networking and Information Technology, Distributed Systems, Java, Real-time distributed

———————————— ◆ ————————————

## 1 INTRODUCTION

LoRaWAN [11] is a low power wide area network. In fact, it is an ultralow power, long range secured network having features like license free spectrum, geolocation etc. already enabled. However, it's an asymmetric network which means that the magnitude of communication to and from the gateways and node devices are not the same. In such a network setup there are always possibilities of packet losses when edge devices transmit data to gateways which subsequently relay them to the network and application server.

In real world IoT use cases there can be two or more sets of such edge devices where the streaming data may need to be joined based on certain keys or partitions and the final output would reflect the overall state when these scenarios are taken in conjunction. Also, exactly once processing mandates the need of a global snapshotting mechanism of the whole system. Explicit acknowledgement is a feasible technique to ensure that data streams are processed only once. However, this is not widely acceptable in a real-world scenario as its more resource consuming in nature. Thus, although we can allow for some lateness in data and maybe ignore redundant information, on-time processing is still the key factor. One of the benefits of LoRaWAN network is that we have the support of more than one gateway for each edge device and the fact that they are geolocation enabled. This helps us to allow early or late processing of multiple data streams and even validate cases for message losses..

―――――――――――――

• *Kunal Chowdhury. E-mail:2019ht12471@ wilp.bits-pilani.ac.in*

*Please note that all acknowledgments should be placed at the end of the paper, before the bibliography (**note that corresponding authorship is not noted in affiliation box, but in acknowledgment section**).*

## 2 LORAWAN NETWORK TOPOLOGY

There are primarily four components in a LoRaWAN network :

1. End Nodes
2. Gateway
3. Network Server
4. Application Server

LoRaWAN has three types of device classes and each one of them has its own characteristics

1. Class A – This is the most energy efficient method which all devices implement. Here the end device initiates transmission.
2. Class B – In this case either the end device initiates the transmission, or the network can do the same at the fixed interval
3. Class C-  Here the transmission can be initiated by the network at any time and the edge device can also transmit back.

Window of opportunity for the network to send a message to the end device is present as well. However, the communication is always asymmetric. In case of Class A, once the uplink transmission is triggered from the end device , the network gets two receive windows to send a message to the device. These windows are periodic in case of Class B whereas in case of Class C these are almost always present.

### 2.1 Adaptive Data Rate (ADR)

Adaptive Data Rate [13] [16] is a technique to optimize the data rates while ensuring that messages are still received at the gateways. When using this technique, the network server will indicate to the end device that it should reduce transmission power or increase data rate. End devices which are close to the gateways should use a lower spreading factor and higher data rate whereas de-vices further away should use a high spreading factor . This algorithm is based on the Max Signal to Noise Ratio (SNR) , Signal to Noise Data Rate (SNR(DR)) and installation margin of the network[12]. (SNR(DR)) is the required signal to noise ratio to successfully perform the demodulation.

For a given device address SNRmax denotes the maximum of the various SNRs reported by different gateways who received the frame. The frame counter , SNRmax and GtwDiversity is maintained as a separate data structure for each node for last 20 frames. Only the best transmission record is kept if there are more than one transmission with the same frame counter i.e., the one with the highest SNRmax.

For a given device address SNRmax denotes the maximum of the various SNRs reported by different gateways who received the frame. The frame counter , SNRmax and GtwDiversity is maintained as a separate data structure for each node for last 20 frames. Only the best transmission record is kept if there are more than one transmission with the same frame counter i.e., the one with the highest SNRmax.

The formula for SNRmargin is given by

$$SNRmargin = SNRm - SNR(DR) - margin\_db$$

## 3  STREAMS , SNAPSHOTS AND WATERMARKS

Internet of Things (IoT) devices always send streaming data from the end devices and there are instances when we need to join two or more streams. These are manifestations of table joins in Database Systems that are implemented at a stream level. The initiator, orchestrator or coordinator process needs to take a global snapshot of the other processes to get an overall view of the data.

In IoT terminology these processes can be thought of as the edge devices. For a set of common keys across streams we must get the latest and most recent snapshot of the corresponding entity data. These are normally partitioned across windows e.g., tumbling, rolling windows etc.

One can implement Chandy Lamport [3] algorithm to get the global snapshot of the processes. In this approach there is a marker message that is first sent by the initiator process. After it records its own snapshot, it sends this message along all its outgoing channels before sending out any more messages. All messages that follow a marker on a channel have been sent by the sender after it took its snapshot. These messages in the channels act as a distinction between the updates that have been included in the snapshot vis-à-vis the ones that are not. Once a process receives a marker it must record its snapshot immediately.

However, in our current discussion the snapshotting is done centrally by an orchestrator rather than any of the gateway devices. At a stream level the windows are open for processing till a certain amount of time has passed and thereafter a watermark trigger indicates that no more updates can be accommodated in the window [15]. Due to intrinsic pauses in the processes, there can be instances of delayed entries in these windows and depending upon the implementation these can or cannot be accommodated.

Nevertheless , the bottom-line is we shouldn't reprocess the stream data, and this is guaranteed in a system which implements the exactly once processing paradigm. Most systems fallback on external caches, in-memory databases etc. to guarantee the same post the transaction is committed. After the necessary commits are complete, explicit acknowledgement is sent to the producer so that the same message is not resent. These snapshots can be even implemented at an operator level [10]. However explicit commits and acknowledgements in case of IoT edge devices can consume lot of bandwidth and battery power.

## 4  STREAM PROCESSING

With the background of LoraWAN devices, ADR, and other concepts of stream processing that we have discussed so far, we now start our discussion on the strategies of processing data streams with respect to the LoraWAN architecture. Initially we come up with a window size for effective processing of the data and thereafter we build the subsequent use cases of detecting and handling message drops.

### 4.1  ADR Piggybacking

We piggyback on the Adaptive Data Rate mechanism of the contributing networks. We assume that the data rate of edge devices $Dev_1$ and $Dev_2$ are $s_1$ and $s_2$ respectively. Also, we assume the average message size transmitted by the devices are $M_1$ and $M_2$. In an ideal scenario the total data would be transmitted in time $M_1/S_1$ and $M_2/S_2$ respectively. As a conservative estimate we start with the least common multiple (LCM) of $M_1/S_1$ and $M_2/S_2$ as the initial window size. We call this the base composite window (BCW).

1.  We maintain two configurable parameters called desired latency (DL) and window factor (WF). The window factor is used to reduce the window size when the desired latency for the system is met i.e., the current window size has lower latency.

2.  Another parameter called the late entries threshold (LT) denotes the upper limit of the entries that can be included in the current processing window when the watermark has passed.

3.  We keep an eye on the number of late entries for the past few windows and observe if LT is breached. If yes, we fallback to either the last known good value for the window size or the moving weighted average as shown below.

## 4.2 Moving Average

The window size for subsequent processing can be calculated as

$$OptimalWindowSize(n, t) = \frac{\sum_{l=1}^{n} m_{n-l} * w_{n-l}}{\sum_{l=1}^{n} m_{n-l}}$$

$$m_{n-l} = \begin{cases} \text{number of messages in (n-t)-th window,} & \text{if LT is not breached,} \\ 0, & \text{otherwise.} \end{cases} \quad (0.1)$$

$$w_{n-l} = \text{size of (n-t)-th window} \quad (0.2)$$

Fig 1 Moving Average formula for determining window size

## 4.3 Pseudo Code

**Algorithm .1:** GETINITIALWINDOWPERIOD($Dev_1, Dev_2, MsgSize_1, MsgSize_2$)

**comment:** Compute the initial window period using device keys and message size

$S1 \leftarrow$ GETADRBASEDDATERATE($Dev_1$)
$S2 \leftarrow$ GETADRBASEDDATERATE($Dev_2$)
$WindowSize \leftarrow$ LCM($S2, S3$)
**return** ($WindowSize$)

**Algorithm .2:** ONRECOMPUTEWINDOWSIZETRIGGER($StreamKey$)

**comment:** Event Listener to recompute window size

$StreamDetails \leftarrow$ GETSTREAMDETAILS($StreamKey$)
$StreamPerfInfo \leftarrow$ GETAVGPERFSTATS($StreamKey$)
**if** $StreamPerfInfo \leftarrow$ GETCURRENTLATENCY() < GETTARGETLATENCY()
 **then**
  $NewWindowPeriod \leftarrow StreamDetails \cdot$ GETCURRENTWINDOWPERIOD() * GETSHRINKINGFACTOR()
  $StreamDetails \cdot$ UPDATEPERIOD($NewWindowPeriod$)
  $StreamDetails \cdot$ SETLASTKNOWNGOODPERIOD($StreamDetails \cdot$ GETCURRENTWINDOWPERIOD())

 **else** GETLATEENTRIESTHRESHOLD() < $StreamDetails \cdot$ GETNUMOFLATEENTRIES()
  $StreamDetails \cdot$ UPDATEPERIOD($StreamDetails \cdot$ GETLASTKNOWNGOODPERIOD())
**return** (1)

Fig 2 Pseudo-code to incorporate the shrinking of windows

## 5 HANDLING AND DETECTING DROPPED MESSAGES

The orchestrator or the process responsible for taking the snapshots will choose a primary gateway for each device. In addition, there is a secondary gateway that will use only supplementary information wherever needed. This is particularly important if we have distributed orchestrators.

### 5.1 Detection

Detection of late or missed messages can be done in three ways –

- **Late events** - When windows are closed for modification, we can detect the entries that are late from the corresponding gateways.
- **Sequence number stamping** - Each message sent by the edge device would be stamped with the current or next sequence number. In case of message drops the orchestrator will detect a mismatch between the current and expected message sequence number. As shown below the expected sequence number should increment by one and in case there is a message loss the orchestrator will process the message until the watermark has been triggered and the sequence numbers are not conflicting. The normal processing of the messages will update the operator state store. However, in any other case when an anomaly is detected we need to handle the scenario appropriately. Here seq number n+1 is lost.

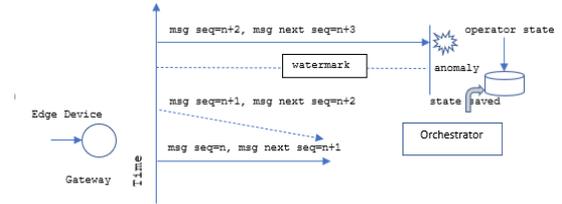

Fig 3 Detecting dropped message using sequence number

- **Event broadcasting** – As mentioned above the orchestrator can designate secondary gateways for edge devices. In this method edge devices will send the message sequence number to the secondary gateways as well. Although the orchestrator only processes the message content from the primary it can reconcile any mismatch of sequence numbers by looking up the sequence vectors in the secondary gateway. Each gateway will thus transmit the sequence numbers of other edge devices as well. The orchestrator will then lookup the sequence number from these gateways and decide whether a message drop has happened. Here orchestrator detects that MsgSeq$_2$ was dropped.

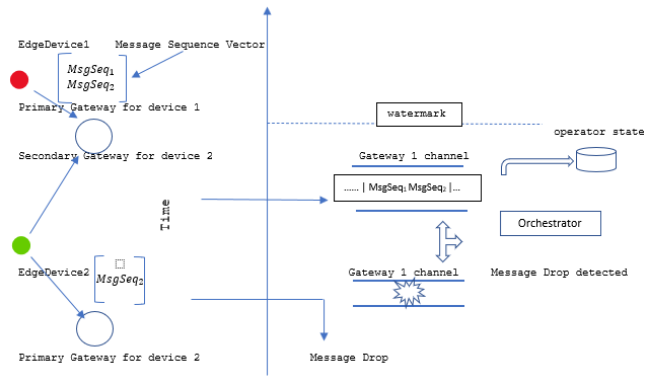

Fig 4 Detecting dropped message using broadcasting

### 5.2 Handling

In the above discussion we investigated two ways to detect the missing messages. Now we present a way to handle the above scenario.

For the following discussion we can refer to the diagram shown below.

1. There can be cases when we are missing data from a particular node, and this is leading to late events in the watermarking process

2.  We assign an odd number of gateways to be connected to each node . e.g., $N_1$ is connected to Gateway$_1$, Gateway$_2$ and Gateway$_3$ to achieve a good quorum.

3.  Each of the gateways can geolocate the nodes and communicate via Class A or Class B mechanism.

4.  We send a message to $N_1$ via Gateway$_1$, Gateway$_2$ using correlations ids  Corr$_1$, Corr$_2$ and ordered by timestamps $t_1$, $t_2$ respectively.

5.  The incoming channels then wait for receiving the response. The snapshotting orchestrator can then relate the response and the corresponding gateway as they are consumed from the queue.

6.  Similarly, the broker then again sends messages via $G_2$ and $G_3$ respectively and then via $G_1$ and $G_3$.

7.  The gateway which relays the maximum number of messages within the stipulated watermarking time will be designated as the primary gateway for the node.

8.  However, we will also keep a backup snapshotting process in case one of the gateways is not behaving correctly.

9.  Once the snapshot is complete it will be written to a persistent store so that we do not duplicate the corresponding state again and maintain the exactly once processing semantics.

10. This approach is applicable for all the contributing streams and in case there is a delay  and watermark must be processed, the old updates from previous window can be sent but they will be eventually rejected down the pipeline.  This can be trivially detected by referring to the id in the operator state store

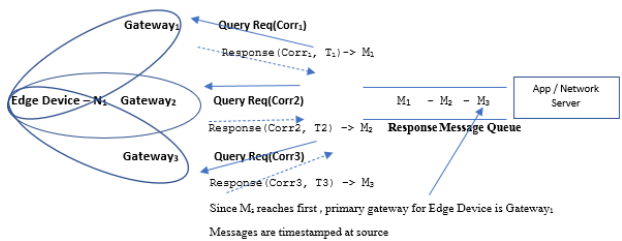

Fig 5 Handling Dropped Message

## 5.3   Early and Late Firing using Geolocation

Based on the geolocation of the edge devices we may allow the early firing of some windows to materialize the processing.  The distance of the edge devices should be measured based on the actual location of the application or network server.   However, as we are joining two streams we need to ensure that the distances for both the devices are greater than a threshold. In such a case we should allow late firing whereas early firing is possible only when both the devices are closer to the gateway

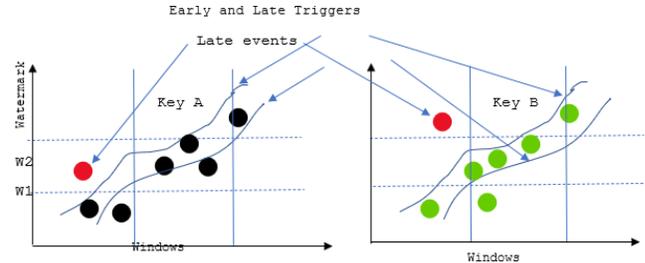

Fig 6 Late and early triggers and late events

## 6   STREAM JOINS

We now delve into some implementation level details of joining the streams by considering the discussions we have done above. In a simplistic form we can assume that the data transmitted from the gateways will be timestamped and stored once it reaches the processing layer. It can be stored in a time series database to which listeners are attached to various accumulator partitions.

An accumulator is an instance of an object that  joins various streams of data, and it will maintain a minimum and maximum allowed timestamp. For scalability we maintain shards or partitions of such accumulators for a set of gateways. Since data across gateways must be joined to process the edge device updates, a secondary merging process is invoked in the pipeline e.g.. the red dot denotes an edge device with some entity updates which are being processed separately via the other gateways as well.

This process will have the facility to detect dropped messages by implementing the strategy discussed above. In such a case the orchestrator will fall back to the method of designating primary and secondary gateways and in the meantime will either allow late entries or republish the stale entries from the state store that will be discarded eventually. In a positive case scenario or once the missing message scenario is handled, operator state will be updated, and the streams will produce partial updates which will be eventually merged for matching keys.

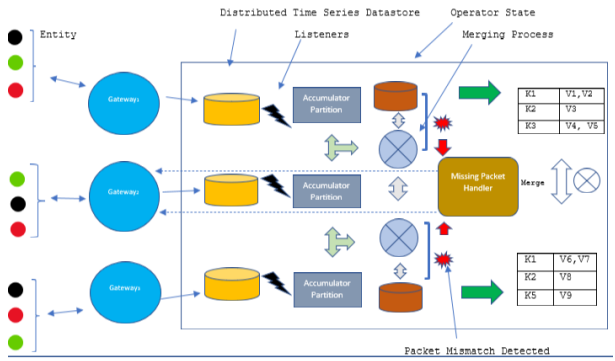

Fig 7 Stream Joins with missing message handler

## 6.1 Class Diagram

Following class diagram depicts a high-level overview of the interfaces (in Java) relating the discussion we had until now. Most of the logic can be implemented in the application server. The operator state is a persistent store of the LDatasource that comes into action during checkpointing by the emitter . Most of the heavy lifting will be done by the gateway pipeline and the accumulator. Stream merging will be done across various accumulators. Generic entities are of two types- BaseEntity and BusinessObject for handling data streams from gateway and objects in the business layer respectively.

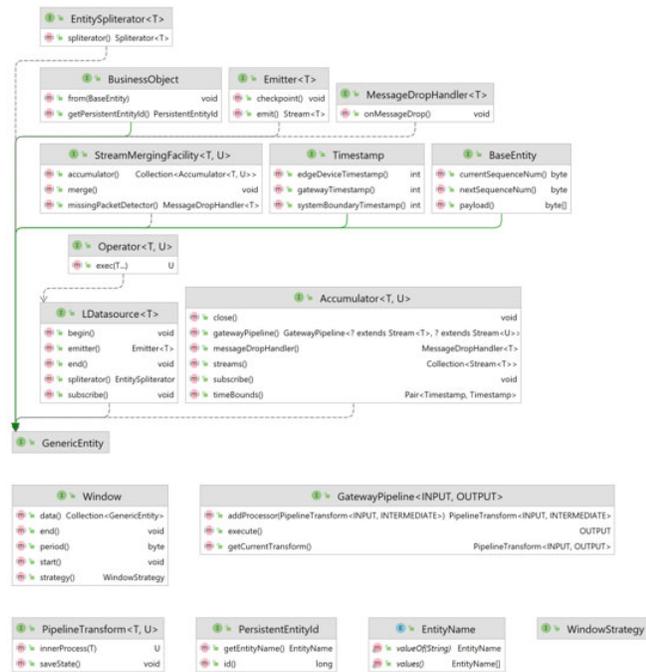

Fig 8 Class Diagram

## 7 ABBREVIATIONS AND ACRONYMS

| ADR | Adaptive Date Rate |
|-----|-------------------|
| DL | Desired Latency |
| IoT | Internet of Things |
| LT | Late Entries Threshold |
| LoRaTM | Long Range |
| RSSI | Received Signal Strength Indicator |
| SF | Spreading Factor |
| SNR | Signal to Noise ratio |
| WAN | Wide Area Network |
| WF | Window Factor |

## 8 CONCLUSION

Here we have discussed a few strategies related to windowing of streaming data, detecting, and handling dropped messages and thereafter connecting all these components for joining data streams in IoT devices. Future work can be on implementations of streaming platforms for LoraWAN devices which are on the move. Some other aspects of these streaming systems can be also from the perspective of reliability and robustness of the end-to-end processing. From a modern technology standpoint there are various frameworks that enable the overall backend design including design of cloud architecture. These aspects can as well be some future improvisations that can be done based on the current work.